# Evaluation of the infectivity of porcine epidemic diarrhea virus on swine vehicles after cleaning and disinfection

Taylor B. Parker, Michael C. Rahe, Kelly A. Meiklejohn, Bradford Sean Darrow, Jason A. Galvis, Gustavo Machado, Juliana Bonin Ferreira*

[1]Department of Population Health and Pathobiology, North Carolina State University, 1060 William Moore Drive, Raleigh, NC, 27606 USA

tbparke2@ncsu.edu
mrahe@ncsu.edu
kameikle@ncsu.edu
bsdarrow@ncsu.edu
joardila@ncsu.edu
gmachad@ncsu.edu
jboninf@ncsu.edu

*designates corresponding author

**Abstract**

Biosecurity measures enforced by the swine industry are in place not only to protect a single farm, but also the vast community of interconnected swine farms. Vehicles play a role in disease movement and given that most pigs move throughout the system multiple times during their lives, vehicle cleanliness is key to preventing disease spread. The objective of this study was to determine the infectivity of PEDV on various swine industry vehicles and evaluate the associated cleaning and disinfection (C&D) methods. Viral swabs were collected from various locations on swine industry vehicles and were used to make the bioassay inoculums which were given to a total of 54 piglets enrolled across two bioassays: a pilot bioassay to ensure the samples were collected and stored in a way that preserved the virus, and a full bioassay testing samples collected from vehicles post C&D. The piglets enrolled in the pilot bioassay tested positive for PEDV within 48 hours, confirming the methods were appropriate and capable of infecting piglets with PEDV. From the full bioassay, the piglets inoculated with samples collected from inside the pigs-to-market vehicle cabins experienced severe diarrhea and tested positive for PEDV. No other experimental treatment group experienced persistent clinical symptoms of PEDV in the full bioassay, suggesting the concentration of infectious PEDV remaining after C&D was not enough to cause severe PED in healthy piglets. However, the cleaning efficacy varied greatly across vehicle types, suggesting that producers should continue to uphold and improve vehicle C&D.

1. **Introduction**

Biosecurity in the swine industry takes many different forms – from boot washes to enforcing downtime between farm visits to cleaning and disinfecting trucks – every farm implements methods to maintain healthy herds. Many of these measures are in place not only to protect a single farm, but also the vast community of swine farms that are interconnected by trucks, personnel, equipment, and the animals themselves (Galvis et al., 2024; Lee et al., 2019). These ideas support the two branches of biosecurity, biocontainment and bioexclusion, which focus on preventing the spread of disease from contaminated herds to healthy herds (Gebreyohannes, 2015). Biocontainment and bioexclusion rely heavily on decontamination protocols, typically consisting of cleaning and disinfecting (C&D). C&D is the removal of pathogens from items that move between farms, including equipment and vehicles (Guan et al., 2017). It is widely known that vehicles play a role in disease movement, and since most pigs in the industry will be moved throughout the system three to four times in their life, vehicle cleanliness is key to preventing diseases from entering new areas (Lee et al., 2019; Lowe et al., 2014; Boniotti et al., 2018).

The swine industry uses a variety of vehicle C&D methods. For trailers that carry live animals, the process starts with the removal of feces, typically via pressure washing. Then, disinfectants are applied, or the trailer may be subjected to high temperatures in a "baker" to further disrupt any pathogens (Houston et al. 2024). Vehicles carrying feed are disinfected regularly, though they aren't always pressure-washed based on staff and resource constraints, and vehicles carrying employees are typically driven through a company carwash. To confirm if a vehicle has been effectively decontaminated, producers commonly collect a representative sample from the vehicle and subject it to reverse transcription-quantitative PCR (RT-qPCR) to detect pathogen RNA. However, a major limitation of RT-qPCR is that it cannot distinguish between non-infectious residual RNA and RNA from live virions. RNA can survive on surfaces much longer than a virus can stay alive outside of a host, so when RT-qPCR is used to detect RNA, the results are not reflective of the infectivity status. Methods such as growing the virus in either cell culture or animal models are the gold standard for studying infectivity.

In a previous study, we evaluated the RNA contamination of porcine epidemic diarrhea virus (PEDV) on swine industry vehicles and found that the C&D methods evaluated, including pressure washing and the application of disinfectants, were mostly ineffective at lowering the concentration of PEDV RNA (Parker et al., 2025). The methods and results described in this manuscript were a continuation of this in which swine bioassays were used to address the following objectives: 1) determine the infectivity of residual PEDV on swine industry vehicles after C&D, and 2) evaluate the efficacy of C&D methods used to decontaminate different vehicle types for disrupting PEDV.

## 2. Materials and Methods

The protocols and sample collections described in this manuscript were approved on December 7th, 2023, and August 21st, 2024, under the North Carolina State University (NCSU) IACUC Protocols 23-441 and 24-326, respectively. This study encompasses two bioassays: 1) a pilot bioassay, conducted in July 2024, and 2) a full bioassay, conducted in March 2025 (**Figure 1**). A total of 54 piglets were enrolled in 18 treatment groups across the two bioassays.

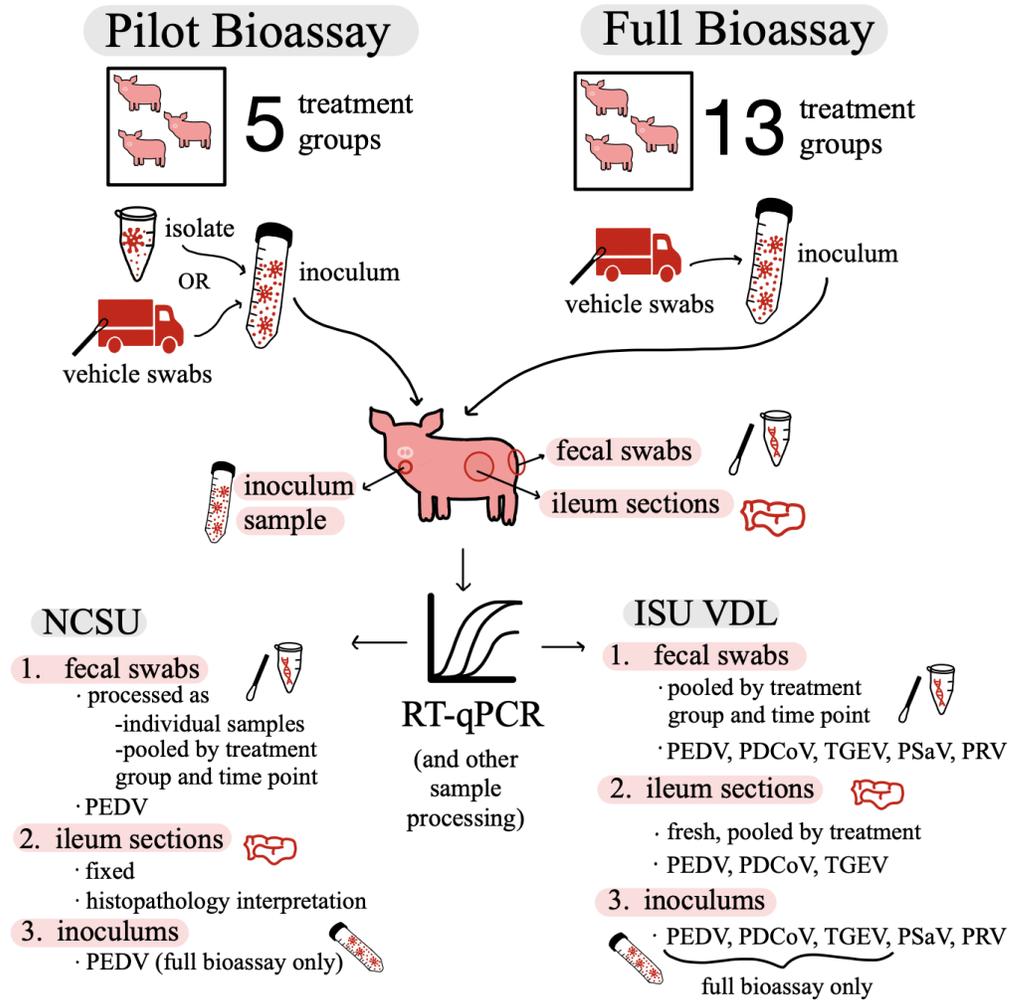

**Figure 1.** Schematic of the methods and sampling workflow used in this study. Abbreviations are as follows: PEDV, Porcine Epidemic Diarrhea Virus; PDCoV, Porcine Delta Coronavirus; TGEV, Transmissible Gastroenteritis Virus; PSaV, Porcine Sapovirus; PRV, Porcine Rotavirus.

*2.1. Swine Bioassay Set-Up*

All piglets were sourced from NCSU's Swine Education Unit (SEU) and were confirmed negative for PEDV, Porcine Deltacoronavirus (PDCoV), Transmissible gastroenteritis virus (TGEV), Porcine sapovirus (PSaV), and Porcine rotavirus (PRV) at the start of each bioassay. Bowls of NutraStart Liqui-Wean Pig Milk Replacer (Milk Specialties Global, Eden Prairie,

MN, USA) were placed in farrowing pens at SEU containing piglets designated for this study to pre-expose them to the food used during the bioassays. Five-day-old piglets were transported from the SEU to the NCSU College of Veterinary Medicine's Laboratory Animal Research (LAR) unit. Upon arrival, each piglet's back was marked with an ID number using a permanent marker. Markings were refreshed every 24 hours. Three pigs were assigned to each treatment group and, to promote well-being, were housed together in a single pen (**Figure 2A**). A single large pen (183 cm x 122 cm) was separated into two treatment pens by metal bars and water-resistant PVC board. If a single room had more than one large pen, they were separated by approximately one meter of space (**Figure 2B**). Additionally, for the full bioassay, the 12 pens were spaced across four rooms (**Figure 2C**). Piglets were fed exclusively the milk replacer (henceforth referred to as "milk"). Fresh milk was prepared every 24 hours according to the manufacturer's recommendations and bowls were cleaned and refilled with 510 mL of milk every four hours for the duration of both bioassays. Each treatment group had separate materials including a food bowl, funnel, sponge, and plastic bottle used to measure and dispense milk.

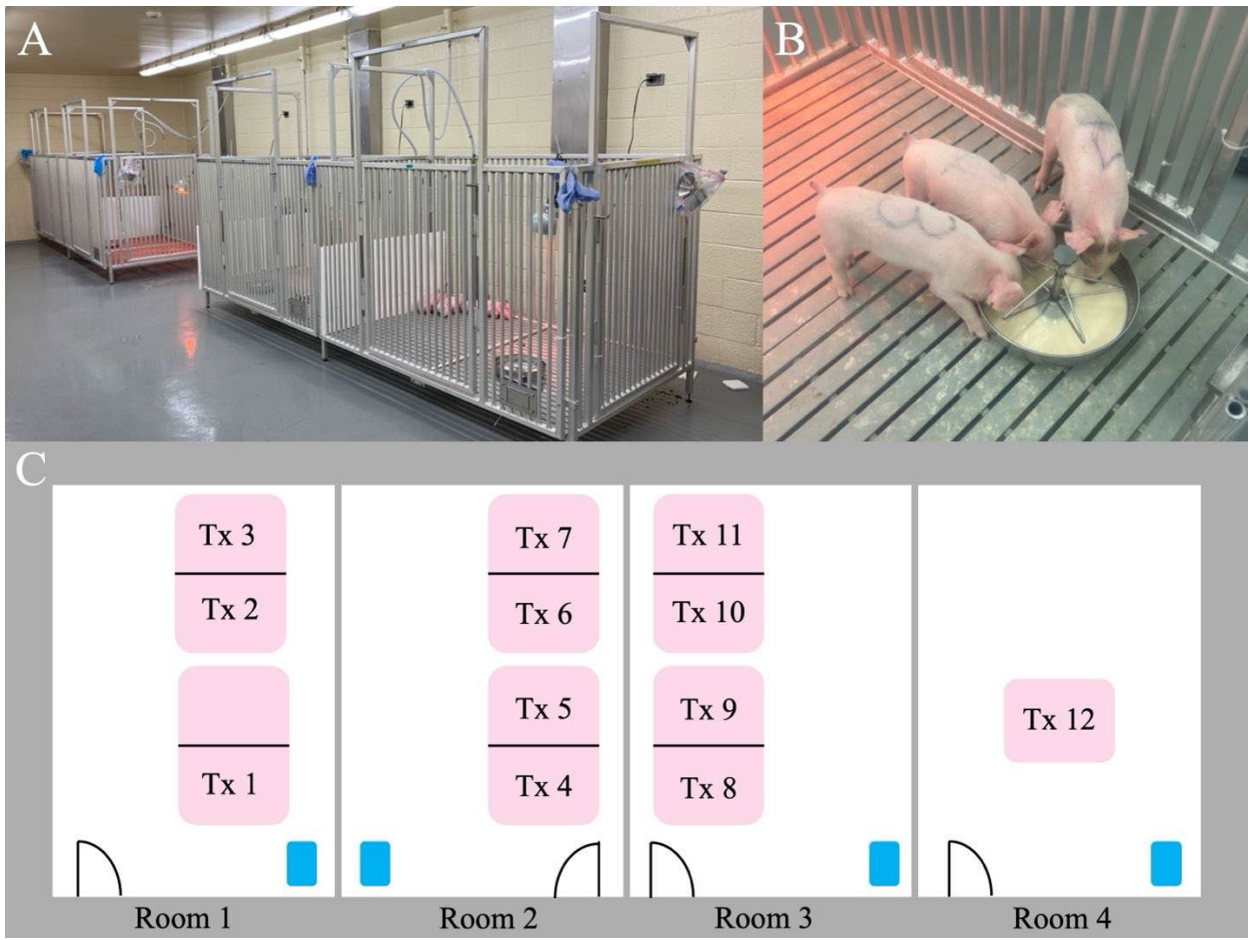

**Figure 2.** Images of bioassay setup. A) Room from pilot bioassay with four pens. White PVC board is seen within a pen to create two treatment groups, with ~1 m of space separating the two large pens. Supplies dedicated to each treatment group can be seen hanging on the sides of pens. B) Three piglets from a single treatment group drinking milk supplement out of a feeder. C) Layout of rooms, pens, and treatment groups used in the full bioassay. Key: dark grey box, NCSU LAR building; White box, individual room within LAR; black outline of a quarter circle, door from antechamber into room; blue box, sink; pink box, pen; black line dividing pink boxes, pen bars/particle board separating two adjoining pens.

*2.2. Treatment group selection*

RNA and viral (VTM) swabs were collected from various locations on swine industry vehicles as described in Parker et al. (2025), including cabins, trailers, and/or tires of a) pigs-to-farm trucks, which were designated by the industry partner as either PED+ or PED- (those with or without a current PEDV outbreak, respectively), b) pigs-to-market trucks (referred to as "market"), c) feed trucks and d) crew trucks. Trucks were disinfected with either a commercially available quaternary ammonium and glutaraldehyde (QAG) or accelerated hydrogen peroxide (AHP) disinfectant. A more in-depth description of the disinfection protocols is provided in Parker et al. (2025). The PEDV RNA concentration obtained from the RNA samples was the basis for determining which treatments likely had the highest and lowest concentrations of PEDV.

*2.2.1. Pilot bioassay*

A pilot bioassay with 15 piglets (eight female and seven male) was conducted to confirm sample collection and storage methods of vehicle samples were adequate for preserving viable virus, as well as to confirm that inoculation methods were sufficient to infect subjects. Piglets were assigned to five treatments as outlined in **Table 1**. Treatment 0 remained with their mother at NCSU's SEU where they had the option of drinking the milk supplement for the duration of the study. Piglets in treatment groups 1-4 were housed at LAR where they were given fresh milk supplement every four hours for the duration of the study. After 72 hours of acclimation, piglets in treatment groups 1-4 were inoculated. PEDV isolate provided by Iowa State University was diluted with VTM and filtered through a 0.22 μM vacuum filter to serve as a positive control in the pilot bioassay (**Table 1**).

**Table 1.** Description of pilot bioassay treatment groups. Each treatment group consisted of three 5-day-old piglets.

| Treatment Group | Inoculum | Inoculum Concentration | Cq of samples used in inoculum* |
|---|---|---|---|
| 0 | None** | - | - |
| 1 | Swine vehicle swabs | Low | ~30-32 |
| 2 | Swine vehicle swabs | High | ~15-26 |
| 3 | PEDV isolate | Low | ~34-36 (Consisted of isolate from Treatment group 4 diluted to $10^8$) |
| 4 | PEDV isolate | High | ~10-12 (TCID50 = $10^4$) |

*Cq determined by in-house PEDV RT-qPCR assay for vehicle swabs and calculated using the TCID50 values provided by ISU for the isolate.
**This negative control was evaluated without an inoculum to ensure the litters selected were healthy, not that the VTM was virus-free. The VTM was filtered with a 0.22 µm filter and prepared in a biological safety cabinet.

### 2.2.2. Full bioassay

Once the pilot bioassay confirmed the vehicle swab collection methods preserved viable virus and inoculation methods were suitable for challenging piglets, a full bioassay was performed to assess whether PEDV RNA-positive samples from vehicles carried live virus after disinfection. Thirteen treatment groups were identified with treatment group 13 remaining at SEU as described in section 2.2.1 for treatment group 0 (**Table 2**). Treatment groups 1-12 were moved to LAR as described in section 2.2.1 with the following modification: piglets enrolled in this bioassay were given 48 hours to acclimatize. This change was made to decrease the amount of overall time the piglets were kept in LAR and was based on the justification that piglets enrolled in the pilot bioassay were consuming milk supplement 24 hours after being moved to LAR.

Table 2. Description of Full Bioassay treatment groups.

| Treatment Group | LAR Room # | Inoculum | Cq of "sister" RNA samples used in inoculum* |
|---|---|---|---|
| 1 | 1 | RT-qPCR Negative (Cq>35) PED+ live haul trailers and cabins | >35 |
| 2 | 1 | PED- live haul trailers disinfected with QAG | 26-35 |
| 3 | 1 | PED- live haul trailers disinfected with AHP | 15-35 |
| 4 | 2 | Feed tires disinfected with QAG | 14-32 |
| 5 | 2 | PED+ live haul trailers disinfected with QAG | 16-32 |
| 6 | 2 | Market live haul trailers disinfected with QAG | 19-29 |
| 7 | 2 | Market live haul trailers disinfected with AHP | 18-25 |
| 8 | 3 | Crew truck tires and cabins (not disinfected) | 18-32 |
| 9 | 3 | Feed and PED- live haul cabins (not disinfected) | 15-28 |
| 10 | 3 | PED+ live haul cabins (not disinfected) | 22-33 |
| 11 | 3 | Market truck cabins (not disinfected) | 23-28 |
| 12 | 4 | Positive control (pre-disinfection samples) | 25-30 |

| 13 | SEU | Negative control (no inoculum) | NA |

*Cq values of the corresponding ("sister") samples used to determine PEDV RNA presence were determined by in-house PEDV RT-qPCR assay (Parker et al., 2025).

*2.3. Inoculum preparation*

A second "sister" swab was collected with every RNA sample collected in Parker et al. (2025). These viral samples were collected in 1 mL of viral transport media (VTM) (Hanks Balanced Salt Solution, 2% fetal bovine serum, 100 µg/mL gentamycin, 0.5 µg/mL amphotericin B, sterilized with 0.22 µm filter) and stored at -80°C. A subset of samples that corresponded to the various bioassay treatments (n= ~30 per treatment) were thawed at room temperature the day of inoculation. The selected viral VTM samples were centrifuged at 12,000 rpm and 4°C for 1 minute to pellet debris. To produce a minimum of 15 mL of inoculum for each treatment group, a micropipette was used to combine up to 1 mL of supernatant from each sample to a 50 mL conical tube for the corresponding treatment group. Conical tubes were centrifuged at 500 rpm and 4°C for 5 minutes. For the pilot bioassay, which consisted of exclusively pre-disinfection samples that had more debris, this step was repeated. Supernatants were poured into a 0.22 µm vacuum filter and inoculum was collected and stored on ice until inoculation.

*2.4. Inoculum delivery and sample preparation*

At the time of inoculation, piglets were removed from the pen one at a time and manually restrained on a surgical table. PEDV inoculum was orally administered via syringe (Tilcare, Valley Point, Singapore) and oral gavage (Valley Vet Supply, Marysville, Kansas, USA) that had been chilled on ice to decrease flexibility and therefore time spent inserting the gavage. For the pilot bioassay, 3 mL were administered per piglet due to constraints on the volume of inoculum available, and for the full bioassay, 5 mL per piglet was used. A new gavage was used for each piglet and new syringes for each treatment. Each piglet received a single dose of antibiotic at 2.5mg/kg (Draxxin, Zoetis, USA) to prevent secondary infection.

After subjects were inoculated, movement between treatment groups was limited to moving from pens that had a lower PEDV RNA concentration to pens that had a higher concentration. Personal protective equipment included hazmat suits, N95 masks, hairnets, socks and rubber boots, plastic boot covers, and gloves. In between pens, gloves were changed and the milk bottles, funnels, and sink area were cleaned with 70% ethanol to reduce the possibility of between-treatment contamination. Feces fell through the pens' slatted flooring onto the concrete floor and were cleaned with water and Virkon (LANXESS Corporation, Pittsburgh, PA, USA) disinfectant as necessary to avoid any feces build-up that contained transmissible virus. Trash was double-bagged and sprayed with Virkon before disposal. Sample boxes were cleaned with Virkon, bagged, sprayed again with Virkon, and bagged again outside of the room.

Rectal swabs were collected in duplicate from each piglet upon arrival to LAR, just before inoculation, and every 24 hours after inoculation for the duration of the study. Additionally, for the full bioassay, swabs were collected 12 hours post-inoculation. Swabs were stored in DNA/RNA Shield (Zymo Research Corporation, Irvine, CA, USA) and kept at -80°C. During swab collection, clinical signs, including diarrhea, vomiting, lameness, and lethargy, were assessed. Once PED symptoms were confirmed, to prevent unnecessary suffering, piglets were humanely euthanized via lethal injection and necropsied.

*2.5. Sample Processing*

For each rectal swab pair, one was submitted to Iowa State University's Veterinary Diagnostic Laboratory (ISU VDL) for RT-qPCR analysis and the other via an in-house RNA isolation and RT-qPCR workflow. The in-house RT-qPCR workflow was validated and utilized in Parker et al. (2025) to minimize cost and maximize the efficiency of detecting PEDV RNA from environmental samples. To maintain consistency, this workflow was utilized to detect PEDV RNA in this study as well. However, since the rectal swabs also needed to be tested for PDCoV, TGEV, PSaV, and PRV, all samples were additionally tested for PEDV at the ISU VDL.

### 2.5.1. In-house sample processing

Rectal swab samples were thawed at room temperature. Swabs were placed in spin baskets mounted in the collection tube (ChromeTech of Wisconsin Inc, Franklin, WI, USA) and centrifuged at 12,000 rpm for two min to draw out liquid from the swab and pellet debris. A total of 375 µL of supernatant was transferred into a clean 1.5 mL tube and either frozen at -80°C or extracted directly after. The Quick-DNA/RNA Viral Kit (Zymo Research Corporation) was used to isolate RNA according to the manufacturer's instructions, with the following modification: Viral DNA/RNA Buffer was left on the sample for 15 min to allow for an adequate lysis period for the high levels of bacteria found in feces. DNA digestion was performed as recommended by the manufacturer to enhance the quality of RNA downstream molecular analyses. Eluates were stored at -80°C. RT-qPCR was performed as described in Parker et al. (2025) with the following exception: accuracy of the in-house assay was evaluated against the results provided by the ISU VDL; therefore, the commercial kit described in the previous study was not used. RT-qPCR results are reported as quantification cycle (Cq) values.

### 2.6. Necropsy and histolopathology interpretation

For the pilot bioassay, all piglets were euthanized at 48 hours post-inoculation (hpi) due to clinical enteric disease in all treatment groups. At necropsy, sections of ileum were collected from each piglet for fixation in 10% neutral buffered formalin. Following fixation, sectioning, tissue processing, and embedding 4 µm formalin fixed paraffin embedded tissue sections were placed on slides for H&E staining. All slides were digitized with a Leica Aperio AT2 slide scanner. Crypt to villus height ratios were collected in ImageScope [v12.4.6.5003] (Leica) and calculated as previously described (Thomas et al., 2015).

For the full bioassay, pigs were euthanized either at 48 hpi due to clinical signs (diarrhea) or at 120 hpi (5 days) per the experimental protocol. At necropsy, fixed sections of ileum and colon were collected with crypt to villus height ratios calculated as described for the pilot bioassay.

**2.7.** *Cleaning and disinfection effectiveness*

We quantified C&D effectiveness by linking the pre-C&D proportion of samples positive for PED via RT-qPCR to the post-C&D bioassay proportion of pigs infected (as determined by the in-house assay only), using a probabilistic model that assumed independent and equal vehicle contributions to infection risk. When letting $N$ be the number of vehicle samples pooled per treatment group, $q$ the proportion of contaminated vehicle samples, $\beta$ the probability that a single contaminated vehicle sample in the pool causes a PEDV-RT-qPCR positive sample in pigs, and $I$ the observed infection probability in the bioassay, we calculated the infection probability using the following formula:

$$I = 1 - (1 - \beta)^{N*q} \qquad (1)$$

Solving for $q$, we estimated the proportion of contaminated vehicle samples:

$$q = \frac{log(1 - I)}{(N * log(1 - \beta))} \text{ where } 0 < \beta < 1 \text{ and } 0 < I < 1 \qquad (2)$$

We calculated $q$ before C&D ($q_1$) as the observed proportion of vehicle samples with viral RNA detected by RT-qPCR, and $q$ after C&D ($q_2$) via Equation 2 (Parker et al., 2025). To ensure that cleaning did not artificially appear harmful, we constrained the inference so that $q_2$ did not exceed the pre-C&D contamination proportion $q_1$ for any treatment group. We set a single global per-vehicle parameter $\beta_i = k/N_i$ that scales with sample pool size to maintain comparability across treatment groups. For each group $i$, we identified the minimum scaling parameter $k_i = N_i\{1 - (1 - I_i)^{1/(N_i q_i)}\}$, and then selected $k = max\ k_i$, which is the smallest global parameter that guarantees $q_{2,i} \leq q_{1,i}$ for all groups. Ultimately, we report effectiveness as a percentage as follows:

$$Effectiveness\ =\ (1 - (\frac{q_2}{q_1}))\ 100 \qquad (3)$$

Along with a sensitivity band for $k$, it ranges from $0.9k$ to $1.2k$.

## 3. Results

*3.1. PEDV prevalence in pilot bioassay*

A total of 48 fecal swabs were collected and submitted to ISU's VDL for PEDV, porcine delta coronavirus (PDCoV), transmissible gastroenteritis virus (TGEV), PRV, and PSaV RT-qPCR testing. Intestinal sections taken from each piglet were combined by treatment group and submitted for PEDV, PDCoV, and TGEV RT-qPCR testing and the inoculums were tested for PRV and PSaV. Five swabs were not collected in duplicate at 48 hpi due to lumen samples having feces present; therefore, 43 fecal swabs were evaluated for PEDV using the in-house RT-qPCR assay described in Parker et al. (2025). PEDV results from both ISU and in-house assays are described in **Table 3**. At 24 hpi, all nine pigs in treatment groups 2, 3, and 4 tested positive for PEDV, were off-feed, and were showing clinical signs of PEDV including diarrhea and vomiting; meanwhile, pigs in treatment group 1 tested negative and were still playful. However, by 48 hpi, pigs in all four treatment groups were experiencing PEDV symptoms and were subsequently and timely euthanized.

**Table 3.** PEDV diagnosis of fecal sample pools, intestines, and inoculums collected from and used in the pilot bioassay as determined by two separate RT-qPCR assays.

| Treatment (Tx) group | RT-qPCR assay | Time of fecal sample (hours post inoculation) | | | Additional sample type | |
| --- | --- | --- | --- | --- | --- | --- |
| | | 0 | 24 | 48 | *Inoculum* | *Intestine* |
| *Tx 1.* Low conc. truck samples | ISU | - | - | + | N/A | + |
| | In-house | - | - | N/A | + | N/A |
| *Tx 2.* High conc. truck samples | ISU | - | + | + | N/A | + |
| | In-house | - | + | N/A | + | N/A |
| *Tx 3.* Low conc. PEDV isolate | ISU | - | + | + | N/A | + |
| | In-house | - | + | + | + | N/A |
| *Tx 4.* High conc. PEDV isolate | ISU | - | + | + | N/A | + |
| | In-house | - | + | + | + | N/A |

Note: Conc refers to concentration. Duplicate samples were not collected for treatments 1 and 2 at 48 hpi and therefore were only processed at ISU. Inoculums were only tested for PEDV using the in-house assay and intestine samples collected after necropsy were only processed at ISU.

3.1.1. *PEDV prevalence and histopathology of intestine*

Intestine samples collected during necropsy from all four treatment groups came back positive for PEDV and negative for PDCoV and TGEV. Additionally, the inoculums (aliquots of the PEDV isolate and the two vehicle sample pools) were tested for PRV and PSaV at ISU and for PEDV using the in-house assay. The PEDV isolate tested negative for PSaV and PRV, whereas the vehicle sample pools tested positive for both. All four inoculums tested positive for PEDV. All fixed sections of ileum from treatment groups 1-4 had lesions of severe atrophic enteritis with low mean villus height/crypt ratios (**Table 4**).

**Table 4.** Pilot bioassay mean villus height/crypt ratio in sections of ileum from three neonatal piglets from each group. Conc refers to concentration.

| Treatment (Tx) group | Villus Height Mean +/- SEM | Crypt Depth Mean +/- SEM | Villus/Crypt ratio Mean +/- SEM |
|---|---|---|---|
| *Tx 1*. Low conc. truck samples | 181.1 +/- 10.66 | 119.4 +/- 17.06 | 1.78 +/- 0.26 |
| *Tx 2*. High conc. truck samples | 199.6 +/- 28.36 | 144.4 +/- 14.75 | 1.51 +/- 0.17 |
| *Tx 3*. Low conc. PEDV isolate | 168.6 +/- 27.70 | 100.9 +/- 6.37 | 1.58 +/- 0.14 |
| *Tx 4*. High conc. PEDV isolate | 190.7 +/- 18.46 | 116.3 +/- 9.73 | 1.76 +/- 0.16 |

3.1.2. *PEDV and other swine virus prevalence in fecal swabs and inoculums*

At 24 hpi PSaV was first detected in treatment group 2. At 48 hpi, treatment groups 1, 2, and 3 had tested positive for PSaV and all treatment groups tested positive for PRV. The PEDV isolate tested negative for both PRV and PSaV, whereas both vehicle sample pools tested positive for both.

*3.2. PEDV prevalence in full bioassay*

For the full bioassay, 270 fecal swabs were collected and submitted to ISU's VDL for PEDV, porcine delta coronavirus (PDCoV), transmissible gastroenteritis virus (TGEV), PRV, and PSaV RT-qPCR testing. A duplicate swab was collected for all swabs except those collected at -48 hpi and processed with the PEDV in-house RT-qPCR assay (n, 234). PEDV results from both ISU and in-house assays are described in **Table 5**.

Thirty-six of the fecal samples submitted to ISU were collected when the piglets first arrived at LAR, 48 hours before they were inoculated (duplicates not collected). All -48 hpi samples were negative for PEDV, PDCoV, TGEV, PRV, and PSaV.

**Table 5.** PEDV diagnosis of fecal sample pools collected from the full bioassay as determined by two separate RT-qPCR assays.

| Treatment (Tx) group | RT-qPCR assay | Time of fecal sample collection (hours post inoculation) | | | | | | |
|---|---|---|---|---|---|---|---|---|
| | | -1 | 12 | 24 | 48 | 72 | 96 | 120 |
| Tx 1; RT-qPCR Negative (Cq>35) PED+ live haul trailers and cabins | ISU | - | - | + | + | - | - | - |
| | In-house | - | - | -* | -* | - | - | - |
| Tx 2; PED- live haul trailers disinfected with QAG | ISU | - | - | - | - | - | - | - |
| | In-house | - | - | - | - | - | - | -* |
| Tx 3; PED- live haul trailers disinfected with AHP | ISU | - | - | - | - | - | - | - |
| | In-house | - | - | - | + | - | - | - |
| Tx 4; feed tires disinfected with QAG | ISU | - | - | - | - | - | - | - |
| | In-house | - | - | - | -* | - | - | - |
| Tx 5; PED+ live haul trailers disinfected with QAG | ISU | - | - | - | - | - | - | - |
| | In-house | - | - | - | + | - | - | - |
| Tx 6; Market live haul trailers disinfected with QAG | ISU | - | - | - | - | - | - | - |
| | In-house | - | - | + | - | - | - | - |
| Tx 7; Market live haul trailers disinfected with AHP | ISU | - | - | + | - | - | - | - |
| | In-house | - | - | - | - | - | - | - |
| Tx 8; Crew truck tires and cabins (not disinfected) | ISU | - | - | - | - | - | - | - |
| | In-house | - | - | - | - | - | - | - |
| Tx 9; Feed and PED- live haul cabins (not disinfected) | ISU | - | - | - | - | - | - | - |
| | In-house | - | - | + | -* | - | - | - |
| Tx 10; PED+ live haul cabins (not disinfected) | ISU | - | - | - | - | - | - | - |
| | In-house | - | - | + | + | - | -* | - |
| *Tx 11*; Market truck cabins (not disinfected) | ISU | - | - | - | - | N/A | N/A | N/A |
| | In-house | - | - | - | + | N/A | N/A | N/A |
| *Tx 12*; Positive control (pre-disinfection samples) | ISU | - | - | + | + | N/A | N/A | N/A |
| | In-house | - | - | + | + | N/A | N/A | N/A |
| *Tx 13***; Negative control (no inoculum) | ISU | N/A | N/A | N/A | N/A | N/A | N/A | N/A |
| | In-house | N/A | N/A | N/A | N/A | N/A | N/A | N/A |

* Represents sample pools where one of the three samples included in the pool tested positive using the in-house RT-qPCR assay, but the pool itself tested negative.

**Fecal swabs were not collected from treatment group 13 due to the risk of cross-contamination and therefore inability to travel between SEU and LAR locations. Piglets were monitored for PEDV symptoms.

### 3.2.1. Clinical signs

Within 24 hours (-24 hpi) of the piglets arriving at LAR, loose stools were observed in treatment groups 3 and 8. One hour before inoculation (-1 hpi), loose stools were still observed in treatment group 8. Samples from both pens were collected at -1 hpi and additionally tested for *Clostridioides difficile* A and B toxins, *Clostridium perfringens*, and *E. coli*. Both pens tested positive for alpha and beta2 strains of *C. perf*, as well as for both *C. diff* A and B toxins. Clinical signs associated with PED began at 24 hpi and are described in **Table 6**. Treatment groups 11 and 12 were euthanized at 48 hpi due to being hyporexic and having severe diarrhea.

**Table 6.** Clinical signs observed during the full bioassay.

| | Time of observation (hours post inoculation) | | | | | | |
|---|---|---|---|---|---|---|---|
| **Treatment (Tx) group** | *-1* | *12* | *24* | *48* | *72* | *96* | *120* |
| *Tx 1*; RT-qPCR Negative (Cq>35) PED+ live haul trailers and cabins | None | None | Vomiting | Diarrhea, vomiting | None | Mild diarrhea | None |

| Treatment | | | | | | | |
|---|---|---|---|---|---|---|---|
| *Tx 2*; PED- live haul trailers disinfected with QAG | None | None | None | None | None | Mild diarrhea | None |
| *Tx 3*; PED- live haul trailers disinfected with AHP | None | None | None | None | None | None | None |
| *Tx 4*; Feed tires disinfected with QAG | None | None | None | None | None | None | None |
| *Tx 5*; PED+ live haul trailers disinfected with QAG | None | None | None | None | Mild diarrhea | None | None |
| *Tx 6*; Market live haul trailers disinfected with QAG | None | None | None | None | None | None | None |
| *Tx 7*; Market live haul trailers disinfected with AHP | None | None | None | None | None | None | None |
| *Tx 8*; Crew truck tires and cabins (not disinfected) | Loose stools | None | None | None | Lethargy | None | None |
| *Tx 9*; Feed and PED- live haul cabins (not disinfected) | None | None | None | None | Lethargy | None | None |
| *Tx 10*; PED+ live haul cabins (not disinfected) | None | None | None | None | Lethargy | None | None |
| *Tx 11*; Market truck cabins (not disinfected) | None | None | Diarrhea, lethargy | Severe diarrhea | N/A | N/A | N/A |
| *Tx 12*; Positive control (pre-disinfection samples) | None | None | Diarrhea, vomiting, lethargy | Severe diarrhea, vomiting, refusal to eat | N/A | N/A | N/A |
| *Tx 13*; Negative control (no inoculum) | None | None | None | None | None | None | None |

### 3.2.2. *PEDV and other swine virus prevalence*

Thirteen intestinal sections were submitted for PEDV (commercial and in-house), PDCoV, and TGEV RT-qPCR tests, and aliquots of the inoculums given to the subjects were tested for PEDV (ISU and in-house), PDCoV, TGEV, PRV and PSaV. No intestines tested positive for PDCoV nor TGEV. Intestines from treatment groups 10 and 12 tested positive for PEDV with the ISU assay, while only treatment group 12 tested positive with the in-house assay; however, the in-house CT for treatment group 10 was only 1.2 above the threshold for determining positive/negative samples. Only treatment group 12 had a villus/crypt ratio compatible with severe atrophic enteritis (**Table 7, Figure 3**; Thomas et al., 2015). Treatment groups 2 and 11 had multifocal small aggregates of neutrophils within the lamina propria at the tips of ileal villi.

**Table 7.** Full bioassay mean villus height/crypt ratio in sections of ileum from three neonatal piglets per group.

| Treatment (Tx) group | Villus Height Mean +/- SEM | Crypt Depth Mean +/- SEM | Villus/Crypt ratio Mean +/- SEM |
|---|---|---|---|
| 1 | 583.54 +/- 29.53 | 162.11 +/- 15.18 | 3.78 +/- 0.52 |
| 2 | 447.09 +/- 5.01 | 129.28 +/- 12.44 | 3.78 +/- 0.21 |
| 3 | 495.68 +/- 28.28 | 137.88 +/- 12.16 | 3.64 +/- 0.03 |
| 4 | 505.34 +/- 73.36 | 145.00 +/- 10.39 | 3.42 +/- 0.81 |
| 5 | 597.69 +/- 94.83 | 132.79 +/- 5.02 | 5.18 +/- 0.78 |
| 6 | 448.09 +/- 67.27 | 135.63 +/- 5.43 | 3.41 +/- 0.40 |
| 7 | 618.98 +/- 72.25 | 130.90 +/- 6.99 | 4.57 +/- 0.73 |
| 8 | 465.38 +/- 28.21 | 129.61 +/- 8.59 | 3.78 +/- 0.52 |
| 9 | 558.06 +/- 80.12 | 132.93 +/- 10.57 | 4.29 +/- 0.57 |
| 10 | 496.07 +/- 32.11 | 131.12 +/- 7.19 | 4.10 +/- 0.11 |
| 11 | 502.79 +/- 27.56 | 139.92 +/- 8.92 | 3.28 +/- 0.72 |
| 12 | 232.08 +/- 33.24 | 148.20 +/- 14.98 | 1.43 +/- 0.63 |

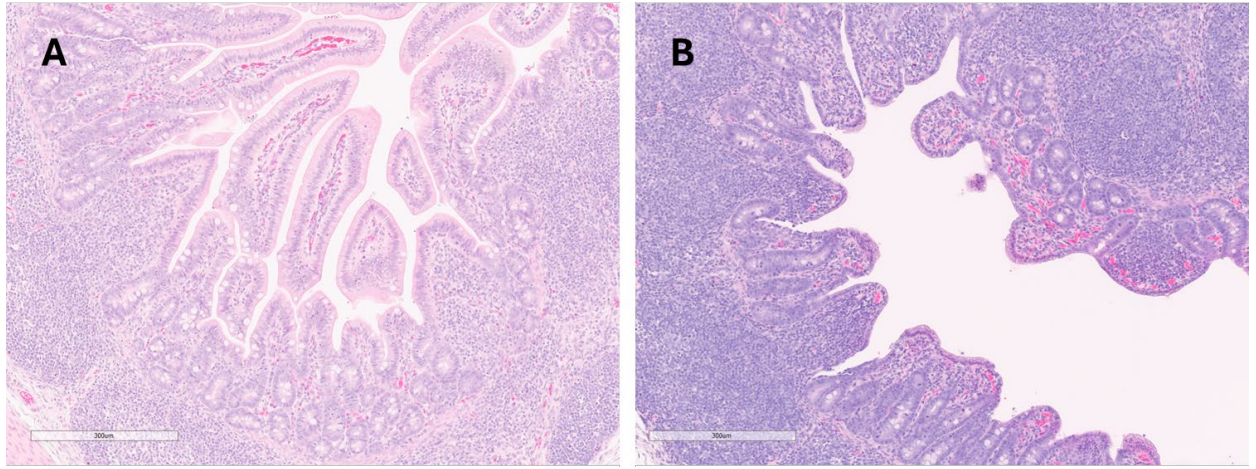

**Figure 3**. Photomicrographs of ileum from piglets with and without PEDV in the full bioassay. A) Treatment Group 3 - unaffected small intestinal mucosa at 120 hours post-inoculation (hpi) (100X magnification). B) Treatment Group 12 - severe atrophic enteritis at 48 hpi (100X magnification).

The only inoculum that tested positive for PEDV with the ISU assay was for treatment group 12, while the only inoculum to test positive for PEDV with the in-house assay was for treatment group 7. When using the in-house assay, PEDV RNA was additionally detected in inoculums for treatment groups 4, 6, 8, 9, 10, and 11, however the Cq for these samples were above the threshold for being considered PEDV positive (Cq values between 33.641 and 36.342). All inoculums were negative for TGEV and only the inoculum for treatment group 1 tested positive for PDCoV. Inoculums from treatment groups 1, 5, and 12 tested positive for PSaV, and all inoculums (treatment groups 1 through 12) tested positive for PRV.

Once a treatment group tested positive for PSaV, they remained positive until the end of the study. Treatment group 1 first tested positive PSaV at 24 hpi, treatment group 5 at 48 hpi, treatment group 10 at 72 hpi, treatment groups 2 and 7 at 96 hpi, and treatment groups 3 and 6 at 120 hpi. Treatment group 11 tested positive for PRV at -1 hpi and at 12 hpi, but subsequently recovered. No other treatment groups tested positive for PRV during the study.

### 3.3. *Cleaning and disinfection effectiveness*

We estimated the effectiveness for each vehicle group (**Table 8**). The scaling parameter $k$ was 2.45, and $\beta$ ranged between 0.08 and 0.1.

**Table 8**. Cleaning and disinfection effectiveness based on a probabilistic model.

| Treatment (Tx) group | $q_1$† | $I$‡ | $q_2$* | Cleaning effectiveness | Sensitivity |
|---|---|---|---|---|---|
| *Tx 2*; PED- live haul trailers with QAG | 0.43 | 0.33 | 0.16 | 63.8% | 59.7% to 70.2% |
| *Tx 3*; PED- live haul trailers with AHP | 0.43 | 0.67 | 0.43 | 0% | -11.6% to 17.4% |
| Tx 4; Feed truck tires with QAG | 0.18 | 0.33 | 0.16 | 12.6% | 2.5% to 27.9% |
| *Tx 5*; PED+ live haul trailers with QAG | 0.71 | 0.67 | 0.43 | 39.5% | 32.4% to 50.1% |
| *Tx 6*; Market trailers with QAG | 0.73 | 0.33 | 0.16 | 78.7% | 76.2% to 82.4% |
| *Tx 7*; Market trailers with AHP | 0.93 | 0 | 0** | 100% | 100% to 100% |

† $q_1$ = Proportion of contaminated vehicle samples before cleaning and disinfection
‡ $I$ = Infection probability
* $q_2$ = Proportion of contaminated vehicle samples after cleaning and disinfection
** Given the cleaning effectiveness was based solely on the numbers of piglets who individually tested PEDV positive via RT-qPCR in a treatment group, results obtained from sample pools were not used. Since no individual pigs tested positive for treatment group 7, the resulting $q_2$ is 0. However, it should be noted that the sample pool from treatment group 7 at 24 hpi did test positive at ISU.

### 4. Discussion

We previously found that the commercial swine vehicle disinfectants quaternary ammonium and glutaraldehyde and advanced hydrogen peroxide were mostly ineffective at lowering the concentration of PEDV RNA (Parker et al., 2025). The work presented here is a continuation of this where we evaluated the infectivity of those same vehicle samples using swine bioassays. PEDV is notoriously challenging to grow in cell culture and is especially rare to successfully propagate from environmental samples (Chen et al., 2014; Oka et al., 2014). Given the difficulties faced when trying to culture environmental samples, we opted to determine infectivity by inoculating piglets with the vehicle samples, which is also considered the gold standard of infectivity models and therefore the most appropriate approach for this study (Puente et al., 2020; Holtkamp et al., 2016).

All fecal swabs in this study were processed with two RT-qPCR assays to detect PEDV RNA: one at ISU and one developed in-house (Parker et al., 2025). The in-house assay was developed to be sensitive enough to detect PEDV in environmental samples while remaining cost-efficient. Given the vehicle RNA samples were processed with this in-house assay, we opted to also process the bioassay samples with this assay, in addition to processing at ISU, for consistency and reliability purposes. For ease of comprehension in the following paragraphs, the results of the two assays used in this study will not be distinguished when discussing the differences in treatment groups.

The first bioassay, referred to as the pilot bioassay, used vehicle samples collected from live-haul trucks carrying pigs from farms deemed PED+ before they were cleaned and disinfected. This confirmed that the methods used for sample collection and storage, and for inoculating piglets, were appropriate and capable of infecting piglets with PEDV. The infectivity of these vehicles before C&D was confirmed when both treatment groups given vehicle samples showed symptoms and tested positive for PEDV within 48 hours.

The full bioassay consisted mostly of treatment groups that encompassed vehicle exteriors that had been through C&D and the interiors of truck cabins that were not disinfected. Additionally, there were three treatment groups with inoculums not from C&D treated vehicles: 1) samples from vehicles moving between PEDV+ farms but the sister RNA swab had a Cq ≥ 35, which would be deemed negative for PEDV using our in-house assay (Parker et al., 2025) (treatment group 1), 2) a positive control group that received pre-C&D samples with sister RNA sample Cq values of 25-31 (all vehicle types and locations) (treatment group 12), and 3) a negative control group which received no inoculum (treatment group 13). The only groups to experience symptoms extreme enough to be euthanized at 48 hpi were treatment groups 11 and 12 (**Table 6**). Due to the volume of samples and time dedicated to feeding the piglets, it was not feasible to process the fecal swabs immediately after collection, and therefore, PEDV status could not be confirmed before euthanasia for the full bioassay. Although only one pig from treatment group 11 tested positive for PEDV, a second pig in the same treatment group had a Cq just 1.57 above the threshold, indicating PEDV was in their system, and it is possible they would have tested positive at 72 hpi. These results, in addition to a) the positive fecal pool result for treatment group 11, b) the PEDV RNA detected in the previous study (Parker et al. 2025), c) the diarrhea observed in the piglets, and d) the lack of co-infections, all support the claim that market vehicle cabins are frequently contaminated with infectious PEDV. The implications of this include drivers potentially spreading live virions when they step out of their cabins and onto farm property (Elijah et al., 2021; Houston et al., 2024; Lowe et al., 2014). Based on our observations, it is common for boots only to be washed at the end of the driver's shift. Ideally, boots should be washed each time the vehicle is washed and boot covers worn every time the driver exits their vehicle.

Although PEDV contamination was the focus of this study, there are many other swine viruses that cause diarrhea in piglets. Testing for these viruses in the bioassays allowed us to evaluate what viruses may be causing the clinical signs observed and whether co-infections were involved. The other treatment groups that were inoculated with cabin interior samples, treatment groups 8, 9, and 10, did not experience anything more than mild lethargy and by the end of the study, no fecal swabs from treatment group 8 tested positive for PEDV (**Tables 5** and **6**). Starting at 72 hpi, treatment group 10 began testing positive for PSaV and not PEDV. However, considering the intestine from this treatment was PEDV positive, these pigs were likely experiencing co-infections, potentially where the PSaV was more virulent (Shi et al., 2021).

Pigs inoculated with crew truck cabin samples (treatment group 8) never produced a positive fecal sample for any of the viruses tested, despite the high levels of PEDV RNA detected inside the cabins of these vehicles in the sister samples (Parker et al., 2025). The employees who use these vehicles are constantly moving throughout the system, which likely increases their chances of picking up viral RNA, but the results of this study indicate that the detected virus is no longer infectious.

As determined in Parker et al. (2025), 88.52% of RNA samples from market trailers post-disinfection tested positive for PEDV RNA. We would expect the $q_2$ values for the market trailer treatments to be similar to this proportion if the disinfectant was not effective against the live virus. However, our $q_2$ values were much lower than expected, but supported by the limited clinical signs of PEDV infection in treatment groups 6 and 7 (**Table 8**). Alternatively, treatments inoculated with the PED+ and PED- trailer post-disinfection samples had $q_2$ values similar to the proportions of positive RNA samples after disinfection (treatment groups 2, 3, and 5; Parker et al., 2025). This suggests that the disinfectants are similarly effective against RNA and live virions when applied to PED+ and PED- trailers.

When comparing treatments inoculated with PED- and market trucks disinfected with QAG (treatment groups 2 and 6) versus AHP (treatment groups 3 and 7), piglets inoculated with AHP disinfected samples experienced less clinical symptoms and tested positive for PSaV later than those inoculated with QAG disinfected samples. Originally, we found that AHP showed decreased effectiveness at disrupting PEDV RNA compared to QAG. Based on the bioassay results, and the resulting cleaning effectiveness values for these treatment groups, AHP is the most effective on market trailers and QAG the most effective on PED- trailers (**Table 8**; Parker et al., 2025; Holtkamp et al., 2016; Bowman et al., 2015).

Treatment group 1 consisted of pre-disinfection samples from the exteriors and interiors of live-haul vehicles coming from PED+ farms, all of whom had sister RNA samples with a PEDV RNA Cq greater than 35. Our assay Cq threshold of 32 was statistically determined to be the equivalent of the Cq threshold of the commercial kit we tested our assay against, so choosing samples with a Cq ≥ 35 should have ensured there was little RNA and therefore little to no infectious virus (Parker et al., 2025). Interestingly, similar to most of the other treatment groups, these piglets had a positive PEDV fecal swab early in the bioassay, experienced mild diarrhea, and were no longer testing positive for PEDV or experiencing clinical signs by the end of the bioassay. These piglets also tested positive for PSaV early in the study and experienced more pronounced symptoms than many of the post-disinfection treatment groups, suggesting that Cq thresholds should be reevaluated.

Limitations associated with the vehicle sample collection are described in Parker et al. (2025). Vehicle samples were stored at -80°C for between 1 and 8 months for the pilot bioassay and 10 and 16 months for the full bioassay. Previous studies using samples that were stored for 11 and 13 months were unsuccessful at producing infectivity (Shumacher et al., 2017; Houston et al., 2024). However, because we had multiple positive tests, this suggests that at least a portion of the samples stored in VTM were still viable after 16 months; inconsistencies in virus detection in the fecal swabs may be a result of some samples no longer being viable at the time of inoculation. The swine virology community would greatly benefit from a standalone study examining PEDV longevity in storage, as previous studies have focused on PEDV stability in the environment (Tun et al., 2016; Kim et al., 2018).

It is important to note that there were multiple discrepancies in the full bioassay samples between the ISU and in-house assays. Of the five PEDV-positive fecal swab pools detected by ISU testing, only two of those pools also tested positive with the in-house assay. This could be because ISU pooled the fecal swabs and then processed, whereas the in-house processing involved extracting individual fecal swabs and subsequently pooling the RNA for RT-qPCR. However, of the three pools that tested negative with the in-house assay, two had one sample from the pool test positive via the in-house assay when processed individually. This suggests that perhaps pooling before extraction versus after extraction impacts PEDV RNA detection. On the other hand, of the nine fecal swab pools that tested positive for PEDV using the in-house assay, only two pools also tested positive with the ISU assay. When originally validating the in-house assay, we found it was more sensitive than the commercial assay it was compared to (Parker et al., 2025). Moreover, there were four instances where neither assay had a positive PEDV result for the pen pool, but there was a singular sample in the pool that did test positive with the in-house assay. These results support the increased sensitivity of the in-house assay. The assays only had matching results for the positive control group, which had Cq values as low as 9.5 for the ISU assay and 6.4 for the in-house assay, suggesting that both assays work the most efficiently when there is a strong singular infection with large quantities of RNA, as opposed to minor co-infections. RT-qPCR-based epidemiology studies are commonly criticized for their lack of applicability when it comes to determining the infectivity of the virus. However, even when a bioassay is implemented, the issue of conflicting RT-qPCR assay results remains an issue. For PEDV, viability RT-qPCR holds the most promising future for surveillance testing as it is capable of determining infectivity without growing the virus and negates the issue of determining Cq cutoff values (Puente et al., 2020; Balestreri et al., 2024).

Our C&D effectiveness calculation approach has several limitations that should be considered when interpreting the results. First, we compared pre-C&D contamination measured by RT-qPCR from vehicle samples with post-C&D infectivity measured by a bioassay, which are fundamentally different. PCR detects viral RNA that does not represent virus infectivity, while bioassays detect infectious outcomes in animals, potentially biasing comparisons. Second, pooling vehicle samples per treatment group in the bioassay and applying a global parameter across all groups likely introduced aggregation bias, which may have obscured group-level heterogeneity in C&D performance. This approach oversimplifies the biological reality, where variation in viral load, sample quality, and vehicle contamination dynamics likely play important roles. Finally, the choice of the per-sample infection probability parameter ($\beta$), defined through a global scaling factor ($k$), introduces uncertainty, as the true probability of infection from a single contaminated vehicle sample remains unknown. Although we included a sensitivity band ($0.9k$ to $1.2k$) to partially account for this uncertainty, it may not fully capture the variability in real-world conditions. Together, these methodological assumptions and simplifications mean that our reported C&D effectiveness values should be interpreted as conservative approximations rather than exact measures of virus inactivation.

Another limitation of this study was the potential for cross-contamination across treatment groups. Due to the biosecurity risks and lack of personnel, we opted not to inoculate our negative control pigs and house them at LAR. For our treatment groups housed at LAR, ideally, all pens would have been in separate rooms or have space between them to reduce this risk. However, since space was limited, we took the following steps to reduce cross-contamination: 1) movement throughout pens was in order from the treatments that were the least likely to be shedding live virus to the treatments that were the most likely, 2) gloves were changed in between working with each treatment group and all PPE was changed when moving between rooms, 3) separate feeding and cleaning equipment was used for each treatment group, and 4) all equipment and sink areas were wiped with 70% ethanol in between working with treatment groups. Despite this being a limitation of the study, pens in close contact with each other are representative of what would occur on a sow farm. A sick litter from one barn can easily spread the virus to the surrounding pens and eventually to other farrowing rooms if piglets or employees are moving throughout the facility without taking proper precautions (Jang et al., 2021).

## 5. Conclusion

The overarching aim of this project was to evaluate decontamination protocols, specifically C&D, used on swine vehicles and their efficacy against both PEDV RNA and infectious virions. Parker et al. (2025) described the PEDV RNA contamination of vehicles both before and after C&D. The study presented here, which used swine bioassays to determine the infectivity of vehicle samples, showed that vehicles coming from PED+ farms do carry infectious levels of PEDV before disinfection, but the concentration of infectious PEDV remaining on these trailers (and other classifications of live haul trailers) after disinfection is not enough to cause severe PED in piglets. Since market vehicle cabins contain enough virus to cause infection, studying disinfection methods of these cabins should be a main priority for researchers. Given the variation in cleaning efficacy across vehicle types, the swine industry should continue to uphold and improve biocontainment methods, including vehicle C&D, and exercise caution when analyzing RT-qPCR results.

**Conflict of interest statement**

All authors confirm that there are no conflicts of interest to declare.

**Ethical statement**

The authors confirm the journal's ethical policies, as noted on the journal's author guidelines page. All animal handling, including protocols and sample collections, were approved under the NCSU IACUC Protocols 23-441 and 24-326 approved on December 7th, 2023 and August 21st, 2024, respectively.

**CRediT authorship contribution statement**

TBP: Data curation, formal analysis, investigation, methodology, project administration, validation, visualization, writing – original draft.

MR: Data curation, investigation, methodology, resources, supervision, validation, writing – original draft.

KAM: Conceptualization, funding acquisition, investigation, methodology, project administration, resources, supervision, validation, writing – original draft.

BSD: Investigation.

JAG: Data curation, formal analysis, software, writing – original draft.

GM: Conceptualization, funding acquisition, software, supervision, visualization, writing – review & editing.

JBF: Conceptualization, data curation, funding acquisition, investigation, methodology, project administration, resources, supervision, validation, writing – original draft.

**Data availability statement**

The data supporting this study's findings are not publicly available and are protected by confidential agreements; therefore, they are not available.

**Funding**

This project is funded by USDA's Animal and Plant Health Inspection Service through the National Animal Disease Preparedness and Response Program via a cooperative agreement between the Animal and Plant Health Inspection Service (APHIS) Veterinary Services (VS) and North Carolina State University, USDA-APHIS Award: AP23VSSP0000C060. The findings and conclusions in this document are those of the author(s) and should not be construed to represent any official USDA or U.S. Government determination or policy.

**Acknowledgements**

The authors would like to thank the production company that allowed us to sample their trucks. We thank Lexus Caster, Laya Kannan S. Alves, Gustavo Donoso Barrera, and Meredith Moss for their assistance with the bioassays. Additionally, for their work on the wet-laboratory processing, we thank Maddy Blong and Karoliina Luik.## References

1. Balestreri, C., Schroeder, D.C., Sampedro, F., Marqués, G., Palowski, A., Urriola, P.E., van de Ligt, J.L.G., Yancy, H.F., Shurson, G.C., 2024. Unexpected thermal stability of two enveloped megaviruses, Emiliania huxleyi virus and African swine fever virus, as measured by viability PCR. Virol J. 21(1), 1. doi: 10.1186/s12985-023-02272-z.

2. Boniotti, M.B., Papetti, A., Bertasio, C., Giacomini, E., Lazzaro, M., Cerioli, M., Faccini, S., Bonilauri, P., Vezzoli, F., Lavazza, A., Alborali, G.L., 2018. Porcine Epidemic Diarrhoea Virus in Italy: Disease spread and the role of transportation. Transbound Emerg Dis. 65(6), 1935–1942. https://doi.org/10.1111/tbed.12974

3. Bowman, A.S., Nolting, J.M., Nelson, S.W., Bliss, N., Stull, J.W., Wang, Q., Premanandan, C., 2015. Effects of disinfection on the molecular detection of porcine epidemic diarrhea virus. Vet microbiol. 179(3-4), 213–218. https://doi.org/10.1016/j.vetmic.2015.05.027

4. Chen, Q., Li, G., Stasko, J., Thomas, J. T., Stensland, W. R., Pillatzki, A. E., Gauger, P. C., Schwartz, K. J., Madson, D., Yoon, K. J., Stevenson, G. W., Burrough, E. R., Harmon, K. M., Main, R. G., Zhang, J., 2014. Isolation and characterization of porcine epidemic diarrhea viruses associated with the 2013 disease outbreak among swine in the United States. J. Clin. Microbiol. 52(1), 234–243. https://doi.org/10.1128/JCM.02820-13

5. Elijah, C., Harrison, O.L., Blomme, Allison, A.K., Woodworth, J.C., Jones, C.K., Paulk, C.B., Gebhardt, J.T., 2022. Understanding the role of feed manufacturing and delivery